\documentclass[journal, onecolumn]{IEEEtran}

\usepackage{cite}

\ifCLASSINFOpdf
\usepackage[pdftex]{graphicx}
\graphicspath{{../pdf/}{../jpeg/}}
\DeclareGraphicsExtensions{.pdf,.jpeg,.png}
\else
\fi
\UseRawInputEncoding
\usepackage{amsmath}
\usepackage{array}
\usepackage{standalone}
\usepackage{etoolbox}
\newtoggle{arXiv}
\newtoggle{bibrun}
\usepackage{graphicx}
\usepackage{color, calc}
\usepackage{array, booktabs}
\usepackage{hyperref}
\usepackage{siunitx, physics}

\usepackage{enumitem}
\setlist[description]{labelindent=0pt, leftmargin=\parindent, font=\normalfont\itshape}
\usepackage{url}
\usepackage{subfigure}
\usepackage{textcomp}
\usepackage{eurosym}
\usepackage{xfrac}
\usepackage{stmaryrd}
\usepackage{makecell}
\usepackage{pdfpages}

\begin{document}

\title{MaRCoS, an open-source electronic control system for low-field MRI}

\author{\IEEEauthorblockN{
    Vlad Negnevitsky\IEEEauthorrefmark{1},
    Yolanda Vives-Gilabert\IEEEauthorrefmark{2},
    Jos\'e M. Algar\'in\IEEEauthorrefmark{3},
    Lincoln Craven-Brightman\IEEEauthorrefmark{4},
    Rub\'en Pellicer-Guridi\IEEEauthorrefmark{5},
    Thomas O'Reilly\IEEEauthorrefmark{6},
    Jason P. Stockmann\IEEEauthorrefmark{4},
    Andrew Webb\IEEEauthorrefmark{6},
    Joseba Alonso\IEEEauthorrefmark{3} and
    Benjamin Menk\"uc\IEEEauthorrefmark{7}
  }

  \small
  \IEEEauthorblockA{\IEEEauthorrefmark{1}Oxford Ionics Ltd, Oxford, OX5 1PF, UK}\\
  \IEEEauthorblockA{\IEEEauthorrefmark{2}Intelligent Data Analysis Laboratory, Department of Electronic Engineering, Universitat de Val\`encia, Valencia, Spain}\\
  \IEEEauthorblockA{\IEEEauthorrefmark{3}MRILab, Institute for Molecular Imaging and Instrumentation (i3M), Spanish National Research Council (CSIC) and Universitat Polit\`ecnica de Val\`encia (UPV), Valencia, Spain}\\
  \IEEEauthorblockA{\IEEEauthorrefmark{4}Massachusetts General Hospital, A. A. Martinos Center for Biomedical Imaging, Charlestown, MA, USA}\\
  \IEEEauthorblockA{\IEEEauthorrefmark{5}Asociaci\'on de investigaci\'on MPC, Donostia-San Sebasti\'an, Spain}\\
  \IEEEauthorblockA{\IEEEauthorrefmark{6}Department of Radiology, Leiden University Medical Center, Leiden, Netherlands}\\
  \IEEEauthorblockA{\IEEEauthorrefmark{7}University of Applied Sciences and Arts Dortmund, Dortmund, Germany}\\

  \thanks{Corresponding author: V. Negnevitsky (vnegnev@gmail.com).}
  \normalsize
}

\date{August 2022}

\maketitle

\begin{abstract}
  Every magnetic resonance imaging (MRI) device requires an electronic control system that handles pulse sequences and signal detection and processing.
  Here we provide details on the architecture and performance of MaRCoS, a MAgnetic Resonance COntrol System developed by an open international community of low-field MRI researchers.
  MaRCoS is inexpensive and can handle cycle-accurate sequences without hard length limitations, rapid bursts of events, and arbitrary waveforms.
  It can also be easily adapted to meet further specifications required by the various academic and private institutions participating in its development.
  We describe the MaRCoS hardware, firmware and software that enable all of the above, including a Python-based graphical user interface for pulse sequence implementation, data processing and image reconstruction.
\end{abstract}

\section{Introduction}

The electronic control system or ``console'' is an indispensable component of any MRI scanner.
This handles the sequences of electromagnetic pulses (both radiofrequency and gradient) in a time-synchronous fashion, as well as the acquisition and digitisation of MR signals, so they can be processed for image reconstruction.
At a higher level, the control system also serves as an interface between the user and the scanner itself.

MRI makes use of pulses in different regimes of the electromagnetic spectrum, radiofrequency in the MHz to hundreds of MHz range, and gradient in the low kHz range, which are combined to fulfill the  requirements for imaging: the creation of phase coherent precessing magnetization using a resonant transmit coil, and spatial encoding, with the encoded signals being detected via Faraday induction by a resonant receiver coil\,\cite{BkHaacke}.
These tasks necessitate a high degree of phase coherence between the various pulses, placing strong constraints on the time control.
Modern field-programmable gate-array (FPGA) modules are perfectly suited for fast, time-synchronous tasks and are thus often at the core of MRI consoles.

The consoles provided by the major scanner manufacturers tend to be tailored to their specific setups\,\cite{Stang2012,Guallart2022}, but there exist also more generic concepts that are somewhat hardware-agnostic and could be used in a wide range of scanners.
Although a number of relatively low cost consoles have been developed for MRI, e.g. Pure Devices GmbH (Rimpar, Germany), Magritek Ltd (Wellington, New Zealand) or Niumag Corporation (Suzhou, China), these inevitably comprise much proprietary hardware and software, and so do not lend themselves to open source development.
In developing an open solution, it is also important to note that many low-field systems actually require quite sophisticated operation to overcome the specific challenges of low-field MRI\,\cite{McDaniel2019,Nakagomi2019,Cooley2020,OReilly2020,Sheth2021,Mazurek2021,Liu2021,Guallart-Naval2022,OReilly2021,Sarracanie2021,Algarin2020,Gonzalez2021,Borreguero2022,Cooley2020b}.

Among the projects opened to the community\,\cite{Stang2012,Tang2015,Hasselwander2016,Anand2018,OCRA,OpenCoreNMR,LimeSDR,Ang2017}, the Open-source Console for Real-time Acquistion (OCRA\,\cite{OCRA}) is notable for its flexibility, inexpensive off-the-shelf components, community focus, and real-time capabilities.
The MAgnetic Resonance COntrol System (MaRCoS\,\cite{Guallart2022}) uses the same versatile hardware, however its software, firmware and FPGA firmware have been replaced to go beyond the limitations of OCRA.

In this article we discuss the MaRCoS rationale and its main performance advances, then present the MaRCoS hardware, firmware and desktop software.
We then discuss the directions in which MaRCoS development is proceeding, and provide some information for readers interested in trying MaRCoS out.


\section{System goals and overview}\label{sec:overview}
MaRCoS was started by a partnership of low-field MRI academic groups\,\cite{Guallart2022} aiming to replace a range of proprietary consoles with a unified, inexpensive yet high-performance platform suitable for some unconventional experiments.

The main project goals were to use open-source or off-the-shelf hardware, be easily programmable, have no hard limitations on the sequence length or complexity, and to be scalable to multiple channels in the future.
Additional goals were to allow the transmit and receive modulation frequencies to be independently varied, and to support sampling rates up to several megahertz for the purpose of investigating highly oversampled data acquisition\,\cite{Galve2020}.
Initially the existing OCRA platform was used and extended\footnote{The authors rewrote the OCRA server and client to use a \texttt{msgpack}-based protocol, support the open-source PulSeq hardware-agnostic sequencing language, and support the newly-developed GPA-FHDO gradient board.}, however its limitations on sequence length, complexity and timing precision, as well as the low-level `assembly'-style sequence programming, led to a complete rewrite of the server, FPGA firmware, and client software to create the MaRCoS system.

\begin{figure}[ht]
\centering
\includegraphics[]{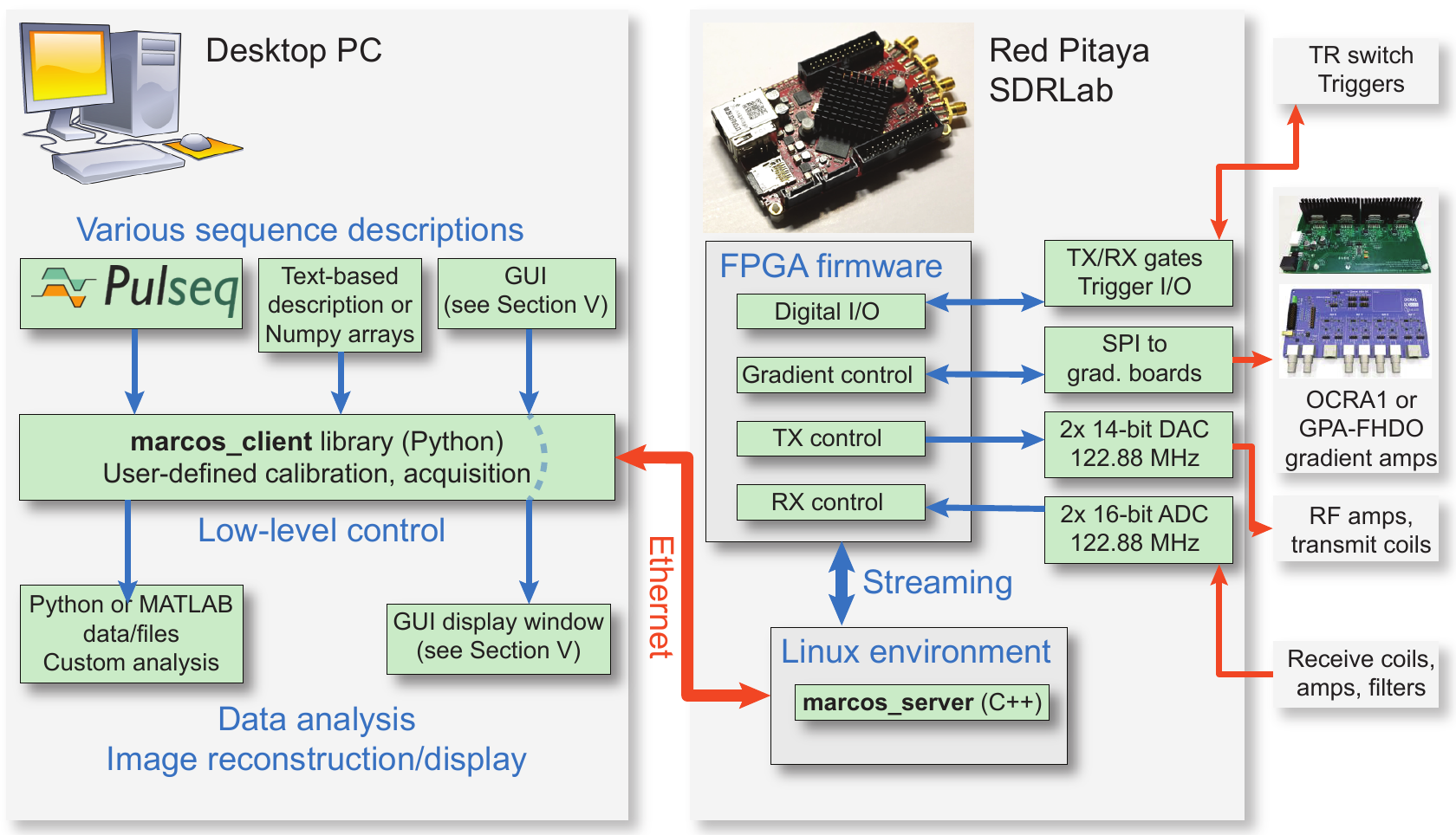}
\caption{ MaRCoS system architecture showing the main components in the desktop software stack and the embedded hardware and software on the SDRLab.
  There are various ways to program a sequence, including a graphical user interface (GUI), direct Numpy arrays and PulSeq, which all control the marcos\_client library.
  This communicates with the SDRLab, executing the sequence and returning the acquired data.
  }
  \label{fig:system_overview}
\end{figure}

The core of MaRCoS is the Red Pitaya SDRLab 122-16\,\cite{sdrlab_website}, a commercial board with two analog inputs for digitising received data and two analog outputs for generating the RF transmit waveforms, with a bandwidth of around 50\,MHz making it suitable for proton MRI at field strengths of up to 1.17 Tesla.
Two receive/transmit channels are run in parallel, with frequency up/down-conversion and the bulk of the filtering handled digitally.
Three digital outputs can be used for controlling RF switches or other peripherals, and one input is used for externally triggering acquisitions\footnote{It will also be used for synchronizing multiple devices during multi-channel operation, which is currently under development.}.
The SDRLab also controls either a GPA-FHDO\,\cite{menkuec_gpa_fhdo_github} or an OCRA1\,\cite{prier_ocra1_overview} four-channel gradient board.
The MaRCoS hardware is controlled from a PC over Ethernet.
\autoref{fig:system_overview} shows the overall system architecture.

On the SDRLab, sequence and acquisition data are streamed to and from an FPGA, allowing for sequences of arbitrary length.
MaRCoS does not use raster clocks or impose any timing constraints on events beyond that of the hardware clock; the architecture permits any change in an internal or external parameter to occur at any time with cycle-accuracy.
This includes changing the RF modulator/demodulator frequencies, RF envelope amplitudes and phases, gradient channel voltages, receiver sampling rates and acquisition timing, and digital outputs.
Sequences are programmed at a low level using simple arrays of times and values for each parameter, which are converted to hardware instructions by the marcos\_client Python library.
There are several intermediate text-based interfaces to suit different styles of programming, as well as a GUI for running a range of calibrations and predefined acquisition routines.

To our knowledge MaRCoS is the first inexpensive system that is capable of handling unlimited, cycle-accurate sequences, handling rapid bursts of events, creating arbitrary waveforms, treating RF and gradients in a uniform way, independently controlling the frequency, phase and amplitude of the different TX and RX channels in each TR, and meeting several other specifications that were required for the experiments being planned when MaRCoS was begun.
The hardware, firmware and software are described in more detail below.

\section{MaRCoS hardware}
\subsection{SDRLab-based console}
The core of the SDRLab is a Xilinx Zynq-7020 system-on-chip (SoC), integrating two embedded ARM processors and a field-programmable gate array (FPGA).
MaRCoS runs Linux and a custom C++ server on the processors, which controls the sequencing and digital signal processing (DSP) on the FPGA as well as digital and analog I/O.
The Linux OS handles networking and provides various standard services such as SSH.

The FPGA part of the Zynq communicates with a dual-channel 16-bit analog-digital converter (ADC) and a dual-channel 14-bit digital-analog converter (DAC), which together provide two independent channels for MRI\footnote{ADC: Analog Devices LTC2185. DAC: Analog Devices AD9767}.
The system is clocked from a 122.88\,MHz crystal oscillator\footnote{Abracon ABLNO-122.880MHz}, providing an analog baseband bandwidth of $\sim 50$\,MHz.
The Zynq also controls multiple digital outputs and inputs, which communicate with the gradient board and external devices such as TR switches.
\begin{figure}[ht]
\centering
\includegraphics[]{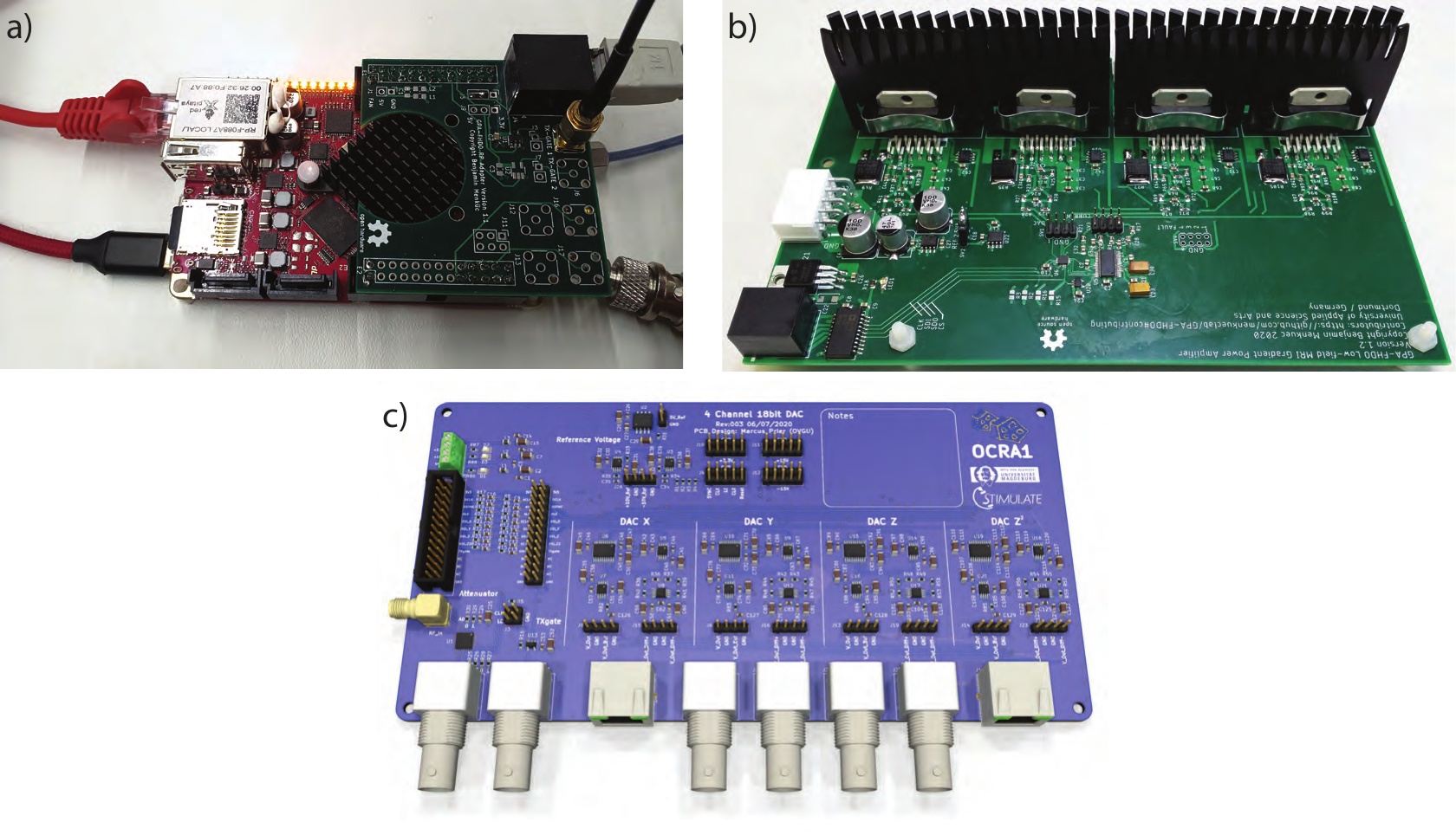}
\caption{MaRCoS hardware components.
  \textbf{a)} Red Pitaya SDRLab with a GPA-FHDO adapter PCB, \textbf{b)} GPA-FHDO board, and \textbf{c)} rendering of the OCRA1 board.
  }
  \label{fig:marcos_hardware}
\end{figure}

\subsection{GPA-FHDO and OCRA1 gradient boards}
MaRCoS currently supports two gradient DAC boards, the GPA-FHDO and the OCRA1, shown in \autoref{fig:marcos_hardware}.
The GPA-FHDO is a gradient DAC that also includes a linear power stage that can deliver $\pm10$\,A on four channels.
The 4-channel 16-bit DAC\footnote{Texas Instruments DAC80504} supports serial-peripheral interface (SPI) clocks up to 50\,MHz.
Due to limitations of the SPI isolator\footnote{Analog Devices ADUM4150} on the GPA-FHDO, the maximum SPI clock is around 40\,MHz, which results in an effective sample rate of around 100\,kSPS if the four channels are updated in parallel\footnote{The SPI bandwidth can be used to update the channels unevenly, e.g. one channel at 400\,kSPS while the others are idle.}.
The GPA-FHDO also monitors the current of each channel, which can be used to automatically calibrate non-linearities and offsets in the analog circuitry.
The ADC\footnote{Texas Instruments ADS8684} has a maximum sample rate of 500\,kSPS for each channel.
This is limited by the SPI isolator in the same way as the DAC sample rate, however since the ADC is mainly used for calibrating the system during the prescan, the sample rate of the ADC is not a critical factor.
An adapter PCB for the SDRLab is available to easily connect to the GPA-FHDO.
It includes a buffer for the TX-gate signal and a connector for an optional fan.
A plugin module is also available for the GPA-FHDO to generate a $\pm12$\,V bipolar output, so that external analog gradient amplifiers can be used.
The GPA-FHDO and adapter PCB designs are fully open-source\,\cite{menkuec_gpa_fhdo_github}, and a set of PCBs can be manufactured and assembled for below \$400.

The OCRA1 is a gradient board with $\pm 10$\,V single-ended and $\pm 20$\,V differential voltage outputs on four channels.
The manufacturing files and schematics are available at \cite{prier_ocra1_overview}.
It uses four 18-bit DACs\footnote{Analog Devices AD5781B}.
The OCRA1 DACs can be run at their maximum SPI clock frequency of 35\,MHz, which results in an effective sample rate of $\sim 400$\,kSPS.
The effective sample rate for the OCRA1 is higher than the GPA-FHDO, mainly because the OCRA1 uses four separate SPI interfaces for its DACs rather than a single four-channel DAC.
Unlike the GPA-FHDO, the OCRA1 does not have a power stage.
Hence, it requires external analog gradient power amplifiers, which users can choose according to their needs.
Another difference between the two boards is that the OCRA1 has an on-board RF attenuator that could be used for flip-angle calibration instead of varying the RF amplitude directly from the SDRLab (MaRCoS does not currently support controlling the attenuator).
The OCRA1 also has an onboard TX gate buffer similar to the SDRLab adapter PCB for the GPA-FHDO.

\section{MaRCoS server and FPGA firmware}
\begin{figure}[ht]
\centering
\includegraphics[]{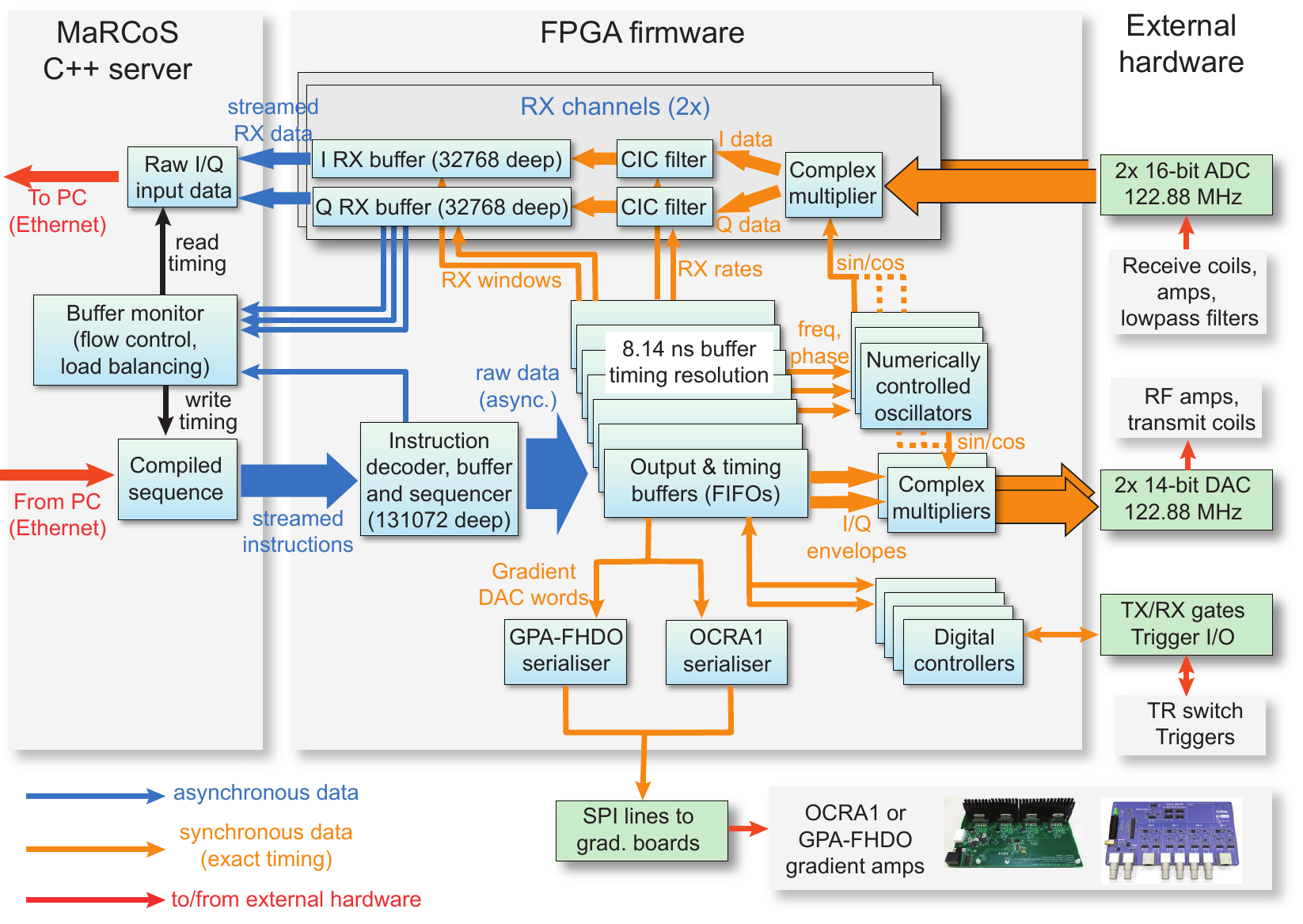}
  \caption{ MaRCoS server and FPGA firmware architecture on the SDRLab.
  The server receives a sequence from the client PC via Ethernet and streams it to the FPGA firmware, where it is translated into time-synchronous hardware operations including RF and gradient outputs.
  The firmware receives data from the ADCs, demodulates and filters it, and saves it into RX buffers, from which it is read by the server and sent to the PC.
  }
  \label{fig:fpga_firmware}
\end{figure}

The MaRCoS architecture was designed from the ground up to achieve the goals in \autoref{sec:overview}.
The FPGA firmware receives instructions from the MaRCoS server and translates them into hardware commands, as well as down-converting, filtering and storing the receive data.
The server in turn receives instruction binaries and control commands from the client PC, and sends back acquired data.
\autoref{fig:fpga_firmware} shows the tightly coupled server-firmware architecture, which relies on the server streaming data to and from the firmware in real time, and the firmware converting between the asynchronous data streams and synchronous (precisely timed) input and output data using several layers of buffering.

\subsection{MaRCoS embedded server}
\label{sec:marcos_server}
The MaRCoS embedded server\footnote{The server is a Linux user-space executable written in C++ running on the SDRLab, and is started/stopped, recompiled, etc. via SSH.} is tightly coupled to the FPGA firmware and has five major roles: receiving sequences from the client PC, controlling the system at a high level (e.g. running or stopping sequences and checking the firmware status), streaming a steady supply of instructions to the FPGA, regularly reading the FPGA receive buffers and streaming the raw input data into large arrays in the system memory, and sending these arrays back to the client PC once the sequence is finished.
When a sequence is executing, the server tries to balance keeping the FPGA instruction buffer near-full against keeping the receive buffers near-empty, since failing to supply enough instructions would result in an unpredictable timing delay, and failing to read the received data would result in a buffer overflow and data loss.
The server continuously tracks the free space in the various buffers and allocates its time based on what is most urgent, and the firmware records 'buffer-full' and 'buffer-empty' events so that the user is always alerted if (and at what point) their sequence overloaded the server.
The server can sustain a steady rate of $\sim 1.5$ million combined reads plus writes per second, which is enough for arbitrary-waveform RF envelopes, gradient outputs and acquisition rates with bandwidths of several hundred kHz.
Bursts of $\sim 20\,000$ output instructions or acquisition samples are possible with several-MHz bandwidths owing to the large buffers\footnote{Once \emph{sustained} rates in the MHz range are required, however, MaRCoS must be modified to support direct-memory access (DMA) in the firmware and server and/or extended with firmware-level data compression.
  This is well beyond current experimental demands.}

The server currently receives a sequence from the client PC, executes it, and returns the data in three separate stages.
Sequences can be hundreds of megabytes in size\footnote{The total size of the MR sequence plus acquired data is limited by the memory that the server can reserve within Linux, which is around 500\,MB.}, however large amounts of sequence and acquisition data can take tens of seconds to transfer to and from the SDRLab.
This could be improved by transferring and executing the sequences in stages, or streaming them from the PC in a similar way to the streaming between the server and FPGA firmware.
The Python-based client software is currently the performance bottleneck, however, so optimising the data transfer is not yet critical.
The client is discussed in \autoref{sec:marcos_client}.

\subsection{FPGA firmware structure}
The FPGA firmware executes the transmit, gradient, digital I/O and control instructions sent from the PC while simultaneously acquiring ADC data, filtering it and storing it in the receive buffers.
The central firmware module is an instruction decoder and sequencer, whose role is to interpret instructions from the circular instruction buffer which is kept filled by the server.
The decoder controls 24 output/timing first-in first-out (FIFO) buffers, which can store 8 words each.
Most instructions tell a particular FIFO to output a data word at a precise time, and these time-synchronous data words control the synchronous firmware modules as shown in \autoref{fig:fpga_firmware}.
These include the various transmit/receive cores, the gradient DAC controllers and digital I/O, as well as cores whose properties are typically controlled asynchronously in other MRI consoles, such as the numerically-controlled oscillator frequency and phase and the cascaded integrator-comb (CIC) filter decimation\footnote{Controlling these with the same timing precision as the RF envelopes and gradients allows in-sequence alterations to the oscillator frequency, acquistion rate, etc. and guarantees that the behaviour of a sequence is reproducible down to the clock cycle.}.
The uniformity of the FIFOs allows each instruction simply to specify a FIFO index, time delay and output word; the complexity of choosing the delays and values is handled entirely in the client software on the PC.
This neatly decouples complex sequence timing and design from low-level hardware or firmware details, since virtually any pattern of FIFO outputs could be executed with no low-level changes.

\subsection{Receive and transmit RF signal processing}\label{sec:rfsignal}
The FPGA firmware performs both the down-conversion and low-pass filtering of the received signal, and the up-conversion of the transmitted signal envelope.
These are traditionally done with external analog components, however can be performed digitally in MaRCoS because the Larmor frequency of LF-MRI systems falls below the 61\,MHz Nyquist frequency of the SDRLab.
On the receive path, the analog signal is digitized by a 16-bit ADC at 122.88\,MHz.
This results in a data rate of 2\,Gbps per channel, which is challenging to transfer from the FPGA to the client PC using inexpensive hardware.

The data rate can be greatly reduced by digitally down-converting, lowpass-filtering and downsampling.
The down-conversion is carried out by multiplying the signal from the ADC by the I and Q outputs of a quadrature oscillator operating at the Larmor frequency.
The numerically-controlled oscillator (NCO) uses a direct digital synthesizer (DDS), whose frequency and phase can be altered during runtime through the output FIFOs.
The firmware has three independent NCOs, and the NCO used by each of the four TX and RX complex multipliers can be independently selected and switched mid-sequence.

In order to avoid aliasing, low-pass filtering is needed before downsampling.
An efficient way to combine decimation and filtering is to use a CIC filter, which consists of a decimator and a moving average low-pass filter.
The decimation rate of the CIC is configured using output FIFOs, in a similar way to the NCO properties.

The aliased noise of a CIC increases towards the edge of the passband.
The frequency response is also not perfectly flat across the passband, thus it is not ideal to use the entire bandwidth of the CIC filter.
Instead, usually a second decimation and filtering stage follows the CIC with a small decimation factor, usually 2-4.
In the case of MaRCoS, the data rate is already greatly reduced after the CIC, and the second decimation stage is not implemented inside the FPGA.
It is good practice to use a finite-impulse-response (FIR) filter as the second decimation stage, which is also tuned to compensate the falling frequency response of the CIC passband.
These filters are called CIC compensation filters.
In order to improve the aliasing performance and stop-band suppression, multiple CIC filters can be cascaded.
The MaRCoS firmware uses a CIC filter with six cascaded stages.

MaRCoS can use either proprietary (Xilinx) or open-source cores for the complex multiplier, NCO and CIC.
The three latter cores were written for MaRCoS, and to the authors' knowledge are among the most feature-complete open-source cores that implement this functionality.
They are available at \cite{menkuec_multiplier, menkuec_cic, menkuec_nco}.

%


%

The transmit chain is simpler than the receive chain, because no filtering is needed.
The complex baseband envelope data, supplied by several output FIFOs, is up-converted by multiplying the signal by the I output of a quadrature oscillator, in a reverse process to the down-conversion.
The real part of the up-converted signal is then sent to the DAC.
A low-pass filter is usually applied in the analog RF chain after the RF power amplifier, in order to remove higher harmonics.


\subsection{Gradient control and digital I/O}
To control the gradient boards, data from the FIFOs is serialized by dedicated modules for the GPA-FHDO and the OCRA1.
The OCRA1 module sends data over four parallel SPI links sharing a clock and enable line, to control the four DACs on the OCRA1 board.
The GPA-FHDO module has a single, bidirectional SPI link, which either controls the four-channel DAC or reads the four-channel current-sensing ADC.
Both modules have adjustable SPI clock frequencies, to optimize the gradient update rate based on the experimental setup and cabling.
They also report errors to the sequencer if they cannot serialize the data they are being provided in time.

The general digital outputs are handled by a single FIFO, and mapped directly to output pins on the SDRLab.
Eight LEDs are controlled in the same way, for debugging and monitoring purposes.
Digital inputs are read at precise times by a special `read' instruction, and a `trigger' instruction can pause the sequencer until a particular digital input changes its state\footnote{This will be used for multi-device synchronization in the future.}.

\subsection{Electronic performance and stability}
The electronic noise levels, amplitude and voltage fluctuations, and timing jitter of MaRCoS should be low enough that they never limit the quality of reconstructed images or cause artefacts/ghosting.

The FPGA firmware is clock cycle-accurate, and residual jitter in the ADCs or DACs comes from the phase noise of the crystal oscillator as well as voltage and temperature fluctuations.
Based on its datasheet\,\cite{oscillator_datasheet} the oscillator has a phase noise of -162\,dBc/Hz at a 10\,kHz offset, and a maximum RMS jitter of 125\,fs over a 12\,kHz bandwidth; this level of jitter is not easily measurable with common laboratory equipment.
A pair of RF pulses 100\,ms apart was sampled using a 1\,GSPS oscilloscope, and the timing jitter between them was below the measurement resolution of 1\,ns.
This measurement is likely too coarse by orders of magnitude to capture the jitter, however it sets an upper bound.

We have also run a phase stability experiment using a CPMG pulse sequence\,\cite{Carr1958,Meiboom1958}, with 100 repetitions of an echo train with 50 echoes, 10\,ms echo spacing and 1\,s repetition time.
The phase of the acquired echoes was stable down to 180\,mrad (standard deviation) for all echoes in the train, and down to $\sim 60$\,mrad for 20\% of them.

The SDRLab ADC and DAC can be characterized by their spurious-free dynamic range (SFDR), which are 90\,dBc and 75\,dBc from their datasheets.
We have not independently measured this, although informal online discussions by SDRLab users indicate the ADC performance is better than the DAC, due to the DAC clock coming from the FPGA whereas the ADC is clocked directly from the crystal oscillator\,\cite{sdrlab_measurements}.

Additionally we have measured the GPA-FHDO voltage drift over several hours to be below 10\,\textmu V, which is well below what would affect typical MRI experiments.

The phase and magnitude of control pulses, as well as their timing, is sufficiently stable for demanding MRI applications\,\cite{Guallart2022}.
Although more thorough characterization is required, the MaRCoS hardware is sufficiently stable for imaging with a range of sequences including gradient echo, turbo spin echo or inversion recovery, with different sampling trajectories like Cartesian or radial\,\cite{Guallart2022}.

\section{Desktop software platform}

\subsection{MaRCoS client library}
\label{sec:marcos_client}
As discussed in \autoref{sec:marcos_server}, the SDRLab executes sequences sent from the client PC, which runs the marcos\_client Python library.
This provides a class for creating and sending sequences from the PC, and various helper routines for controlling the SDRLab hardware, calibrating the GPA-FHDO and testing various aspects of the MaRCoS system.
At the marcos\_client level, a sequence is specified by supplying pairs of (time, value) arrays for each hardware output and control, such as the TX envelope, gradient waveforms, etc.\footnote{For example, to specify two pulses of 70\% full-scale amplitude starting at 20\,\textmu s and 100\,\textmu s from the beginning of the sequence, each with a duration of 30\,\textmu s, one would supply the array pair ([20, 50, 100, 130], [0.7, 0, 0.7, 0]).}.
The library scales and rounds the arrays to machine units, applies time offsets to cancel out latencies in the gradient serialisers and other hardware, and compiles these values and times into binary instructions.
These are encoded via the \texttt{msgpack} protocol and sent to the MaRCoS server, which executes the sequence and returns the acquired data.
The acquired data is then scaled and passed to the user code, where it can be decimated, filtered, then analysed or reconstructed into an image.

The marcos\_client library is the foundation for the higher-level interfaces of MaRCoS, including text-based interfaces such as PulSeq and a graphical interface.
Internally marcos\_client is not yet optimized, and takes tens of seconds to compile when sequences consisting of hundreds of complex TRs are run.
As this becomes a bottleneck, the code can be sped up in a range of ways, or ported to a language such as C++ or Rust for further performance gains -- these will become increasingly important as MaRCoS is scaled to handle more than two TX and RX channels.
As mentioned previously the client-server architecture can also be modified to support streaming sequences, which will further improve performance from the user's perspective.

The MaRCoS server-FPGA firmware stack can be emulated on a PC by using the Verilator tool\,\cite{verilator} to create a clock cycle-accurate C++ model of the FPGA firmware, and compiling the MaRCoS server on desktop Linux so that it directly interacts with the model.
The emulated stack behaves like a virtual SDRLab on the network.
It accepts MaRCoS binary sequences and can produce plots of the emulated hardware outputs, as well as complete debugging information from the FPGA firmware, and is useful for new users to learn the MaRCoS software without requiring any hardware.
The marcos\_client library also has a unit test suite that checks the behaviour of the emulated stack against pre-defined reference outputs, which greatly simplifies development and debugging of the MaRCoS server and FPGA firmware.

\subsection{Text-based interfaces}
We have created a wrapper that enables MaRCoS users to program pulse sequences in the open-source PulSeq language in either Matlab\,\cite{Layton2017} or Python\,\cite{Ravi2019}.
The wrapper converts pulse sequences from the PulSeq text file format (which describes RF pulses, gradient pulses, and other sequence events) into Numpy arrays that marcos\_client takes as inputs, and is available at \cite{FlocraPulseq}.
Advantages of PulSeq are its comparative simplicity and its ability to compile the same script to generate sequences for multiple platforms (including Siemens, GE, and Bruker), helping enable reproducible research across sites.
The MaRCoS console is now part of this PulSeq pulse sequence programming ecosystem.
PulSeq also provides an easy-to-use tool for students in university courses.
Some tools built on top of this include basic scripts for finding the center frequency, calibrating the RF transmit power, calibrating the gradient amplitudes, and basic B0 shimming\,\cite{MGHMARCOS}.

Another MaRCoS wrapper, included with the marcos\_client library, provides a domain-specific language that is similar to Magritek consoles, offering another style of programming.
Due to the flexibility of the array-style format accepted by marcos\_client, new text-based wrappers are easy to write.

\subsection{Graphical User Interface (GUI)}
A GUI was designed and developed to facilitate the interaction between the users and MaRCoS. It is available to download at the PhysioMRI github repository\,\cite{GUI_github}.
We note that the GUI is intended to be used in laboratory environments and not yet for clinical purposes due to lack of required certification.
It offers highly customizable sequences through numerous input parameters that can be modified by the user.

The GUI was designed and developed using Python, which allows for fast application development and execution, and relies on the PyQt5 library.
PyQt5 is a Python wrapper of Qt, which is a widely used GUI toolkit written in C++.
The current version of the GUI has been tested on Ubuntu 20.04.4 LTS and Windows 10, and it is expected to work under MacOS systems.

\begin{figure}[ht]
\includegraphics[]{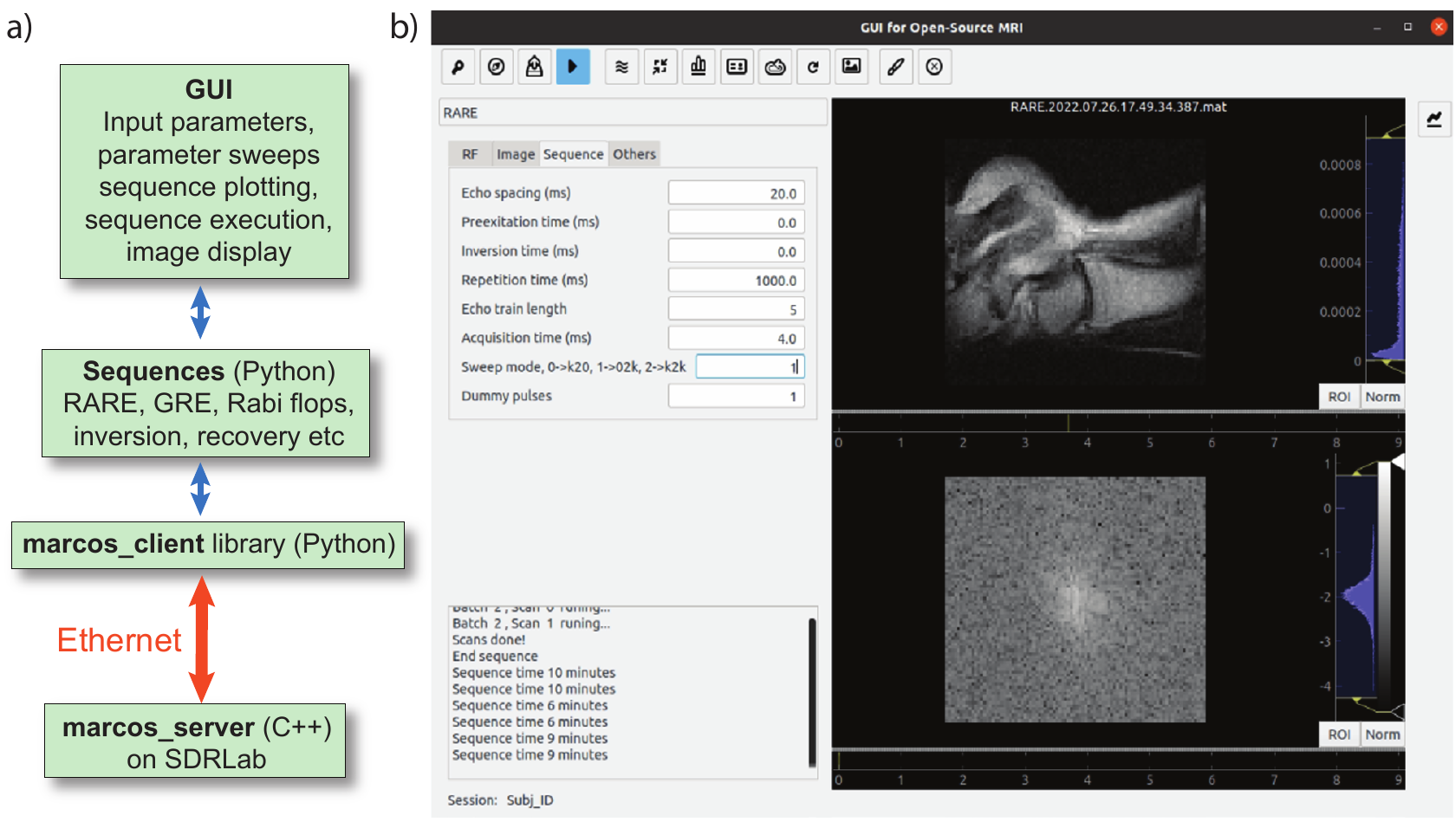}
  \caption{MaRCoS GUI overview. \textbf{a)} Communication and software hierarchy between the GUI and the MaRCoS server on the SDRLab. \textbf{b)} Main window of the GUI.}
  \label{fig:gui}
\end{figure}

When the GUI starts, the latest FPGA firmware and MaRCoS server are installed onto the SDRLab.
The MaRCoS server is immediately run afterwards and the first window of the GUI appears, called the \textit{Session Window}.
The \textit{Session Window} allows the user to select the kind of element that is going to be scanned, which can be a human subject or phantom, and to introduce an ID related to it.
Other information can also be introduced for subjects and phantoms, i.e. demographic data such as sex, height and weight in the case of subjects.
The acquisitions made in the subsequent windows will be related to this session ID: however, the user can change it at any time by clicking the corresponding icon to come back to the session window.
After that, the GUI main window is launched (\autoref{fig:gui}).

The GUI main window was developed combining two methodologies: 1) graphically by manually placing the different layouts in the window by means of Qt Designer\,\cite{qtdesigner}, which generates a ui file; and 2) hand-coding the dynamic addition of elements and widgets specific to the selected sequence into the previously defined layouts.

The communication between the GUI and the marcos\_client library is performed through the sequences, also written in Python, which act as intermediate layer  (\autoref{fig:gui}).
This intermediate layer provides some advantages: 1) the GUI can be safely modified or even substituted by a new one while preserving sequences unchanged; and 2) users can run sequences even without executing the GUI, especially useful while debugging new sequences.
Experiments are run with a predefined 6-fold oversampling factor applied to the acquisition bandwidth required by the sequence.
Once the acquired data is received, the decimation FIR filter implemented in Scipy is applied to extract the signal with the originally required bandwidth.
This oversampling factor compensates the non-flat frequency response of the CIC filter applied in the FPGA, as explained above in \autoref{sec:rfsignal}.

A number of sequences have already been developed and introduced into the GUI repository, which are\,\cite{HandbookPulses}: 3D turbo spin echo, 3D gradient echo, and 2D HASTE, while new sequences can be integrated very easily.
In order to facilitate this, a Python class for each new sequence has to be created.
These sequence classes inherit from a parent class (\texttt{mriBlankSequence}) that contains many methods common to all sequences, e.g. RF pulses, gradient pulses, readout or save data.

It is essential that new sequence classes include three basic elements: the input parameters, their corresponding default values and the associated labels that will be displayed in the GUI main window.
Moreover, they have to include four methods: \texttt{sequenceInfo}, which displays information related to the author of the sequence as well as any useful information related to the sequence itself; \texttt{sequenceTime}, which estimates the duration of the acquisition every time that a GUI input parameter is changed and it prints the result into the console; \texttt{sequenceRun}, which contains the definition of the pulses and timings of the sequence; and \texttt{sequenceAnalysis}, which processes the acquired data and defines the plots that the GUI will display.

The GUI also integrates a calibration window with useful functions that helps to calibrate the scanner.
At the moment, the calibration functions already developed are: Rabi flops for flip angle calibration, inversion recovery sequence to measure $T_1$, resonant  frequency to compensate for any field drift, noise measurement to characterize the system in new environments, a CPMG pulse sequence to measure $T_2$, shimming and generic slice selection.
The calibration window was designed following the same procedure as the main window and its appearance is very similar.

Other functionalities implemented in the GUI main window cover the export of the selected sequence and loading and saving input parameters from and to files.
The export is also useful for the batch acquisition functionality, which performs multiple measurements sequentially without human intervention.
In this case, the user exports the parameters of the different experiments to be performed and uploads them in the desired order to the batch window.
The GUI also includes the possibility to linearly sweep two parameters for any image or calibration sequence.

Finally, the GUI is linked to an XNAT server\,\cite{XNAT}.
XNAT is an open source imaging informatics platform that facilitates common management, productivity, and quality assurance tasks for imaging and associated data.
By enabling the corresponding icon in the GUI main window, it is possible to automatically upload the recently acquired MRIs to the XNAT if it is operational in the PC.
Data is saved in Matlab files (.mat) and NIfTI format.

The GUI is routinely used at i3M.
However, there are a number of developments which we wish to add, such as an automatic method that checks if the sequence parameters are within hardware limits and avoids executing sequences that are out of bounds, the use of sequences that offer the possibility to manage multiple TX or RX channels, and the use of different frequencies for TX and RX.

Additional sequences such as balanced steady state free precession, echo planar imaging or zero echo time, as well as new sampling trajectories beyond Cartesian trajectories, will enrich the possibilities offered by the GUI.
Other useful utilities, which are often featured in commercial imaging systems, include the possibility to set the field of view directly from a scout image, or be able to select a region of interest and show the mean value and standard deviation to get a quick estimation of the signal-to-noise ratio.
Another longer-term aim is to make the GUI accessible to non-expert users, i.e. to develop protocols to generate a standardized set of images from different sequences.

\section{Conclusion}
This paper has described the design and initial performance of MaRCoS, an open-source MRI electronic control system based on a Red Pitaya SDRLab platform and custom gradient boards.
The system has two transmit channels, two receive channels, and can drive three gradient DACs with another channel also available.
The control and testing can be performed in readily available open-source software.
Critical measures of phase stability, timing accuracy, and data transfer rates have been evaluated.
The receiver channel can directly digitize the MR signal, with significant over-sampling enabling the minimization of quantization noise.
CIC filters are used in the final stage of signal processing.
A graphical user interface has been designed in Python, which provides a critical interface in being able to write new imaging sequences and to assess image performance.
The development of the MaRCoS system is just beginning, and a number of future advances and desirable extensions have also been listed.
The low cost means that interested readers can be up-and-running for around \$1000 (for the SDRLab and gradient board).
There are a number of constantly upgraded and interactive resources to get started, including a wiki page\,\cite{MaRCoS} and Slack channel\,\cite{marcos_slack}.



\section*{Author contributions}
J. Alonso, A. Webb and V. Negnevitsky conceived and planned the MaRCoS project, with major technical contributions and advice from B. Menk\"uc and J. P. Stockmann.
B. Menk\"uc wrote the GPA-FHDO serializer, CIC filter, complex multiplier and NCO modules of the MaRCoS firmware, and V. Negnevitsky wrote the rest of the firmware.
L. Craven-Brightman wrote the MaRCoS-PulSeq library.
Y. Vives-Gilabert and J. M. Algar\'in wrote the MaRCoS GUI.
V. Negnevitsky wrote the MaRCoS server, client and auxiliary tools.
B. Menk\"uc, J. M. Algar\'in,  R. Pellicer-Guridi, T. O'Reilly, Y. Vives-Gilabert and L. Craven-Brightman tested and characterized the system, in the research laboratories of J. Alonso, B. Menk\"uc, A. Webb and J. P. Stockmann.
V. Negnevitsky, B. Menk\"uc, J. Alonso, Y. Vives-Gilabert, J. M. Algar\'in, J. P. Stockmann and A. Webb prepared the manuscript, with input from all authors.

\section*{Acknowledgements}
Enormous credit goes to Thomas Witzel, Marcus Prier, Suma Anand and David Schote for their work on the OCRA system as well as many fruitful discussions. Thanks to Danny Park for preparing the original SDRLab Linux image used in both OCRA and MaRCoS, and Maxim Zaitsev for fruitful discussions on rounding and timing in PulSeq.
This work was supported by the Ministerio de Ciencia e Innovaci\'on of Spain through research grant PID2019-111436RB-C21.
The work was co-financed by the European Union through the Programa Operativo del Fondo Europeo de Desarrollo Regional (FEDER) of the Comunitat Valenciana (IDIFEDER/2018/022 and IDIFEDER/2021/004) and ERC Advanced Grant 101021218 PASMAR, the National Institutes of Health NIBIB under grant number U24-EB028984, and the HIFF program of the University of Applied Sciences and Arts Dortmund.

\bibliography{refs}

\end{document}